\documentclass[aip,jcp,reprint,superscriptaddress,citeautoscript]{revtex4-1}
\usepackage[latin1]{inputenc}
\usepackage{graphicx}
\usepackage{fixmath}
\usepackage[colorlinks=true,linkcolor=black,%
citecolor=black,urlcolor=black]{hyperref}

\usepackage{amssymb}
\usepackage{amsmath}
\usepackage{amsthm}
\usepackage{bm}
\usepackage{multirow}
\usepackage{psfrag}
\usepackage{mathrsfs}
\usepackage{dcolumn}
\usepackage[normalem]{ulem} %

\newcommand{\etal}{\emph{et al.}}
\newcommand{\te}[1]{\bm{\underline{#1}}}

\DeclareMathOperator{\erfc}{erfc}
\DeclareMathOperator{\erf}{erf}

\newcommand{\oneline}[5]{
  #1 & #2 & #3 & #4 & #5  \\}

\newcommand{\twoline}[2]{
  #1 & #2  \\}

\newcommand{\itap}{Institut f\"ur Theoretische und Angewandte Physik (ITAP),
  Universit\"at Stuttgart, Pfaffenwaldring 57, 70550 Stuttgart, Germany}
\newcommand{\udem}{D\'{e}partement de Physique and Regroupement
  Qu\'{e}b\'{e}cois sur les Mat\'{e}riaux de Pointe (RQMP),
  Universit\'{e} de Montr\'{e}al, C.P. 6128, Succursale Centre-Ville,
  Montr\'{e}al, Qu\'{e}bec, Canada H3C~3J7}

\begin{document}
\title{Ab initio based polarizable force field generation and application to liquid silica and magnesia}

\date{\today}

\author{Philipp Beck}\email{beck@itap.physik.uni-stuttgart.de}
\affiliation{\itap}
\author{Peter Brommer}
\affiliation{\udem}\affiliation{\itap}
\author{Johannes Roth}
\affiliation{\itap}
\author{Hans-Rainer Trebin}
\affiliation{\itap}

\begin{abstract}
  We extend the program \emph{potfit}, which generates effective
  atomic interaction potentials from \emph{ab initio} data, to
  electrostatic interactions and induced dipoles. 
  The potential parametrization algorithm uses the Wolf direct, pairwise 
  summation method with spherical truncation. The
  polarizability of oxygen atoms is modeled with the 
  Tangney-Scandolo interatomic force field approach.
  Due to the Wolf summation, the computational effort in simulation scales linearly 
  in the number of particles, despite the presence of electrostatic interactions.
  Thus, this model allows to perform large-scale molecular dynamics
  simulations of metal oxides with realistic
  potentials.  Details of the implementation are given, and the
  generation of potentials
  for SiO$_2$ and MgO is demonstrated. The
  approach is validated by simulations of microstructural,
  thermodynamic and vibrational properties of liquid silica and magnesia.
\end{abstract}

\maketitle

\section{Introduction}
\label{sec:intro}

As oxides are important in many technological applications, they have
been object of many experimental 
investigations\cite{Mei2007,Speziale2001,Boiocchi2001} as well
as of theoretical calculations. \cite{itapdb:Karki2007,Vuilleumier2009,Kermode2010,Karki2006,Alfe2005}
Despite continuous advances, \emph{ab initio} methods cannot yet tackle
the study of various material properties due to length and time scale restrictions. 
Classical effective potentials greatly increase the
accessible system sizes and times; the quality of
such simulations, however, depends crucially on the interaction models used.
In this work, we describe a method to derive effective potentials for oxides in
large-scale simulations.

Both force field generation and simulation of oxide systems are
computationally much more demanding than those of metals or covalent
materials due to long-range electrostatic interactions. There have
been many attempts to create effective interaction potentials for
oxides in the last decades. We combine 
two methods that have been applied successfully:
First, in addition to charges we regard
higher moments. It has been shown
for silica\cite{itapdb:Tangney2002, itapdb:Herzbach2005} that polarization effects of the
oxygen atoms have to be taken into account and yield adequate results. 
The potential model of Tangney and Scandolo\cite{itapdb:Tangney2002} (TS) determines the
dipole moments by iteration to a self-consistent solution. This polarizable ion model
yields significantly better results for many properties compared to
other approaches. \cite{itapdb:Herzbach2005}
Second, long-range interactions must be handled correctly
and efficiently in Molecular Dynamics (MD) simulations. There are several methods with the one of
Ewald\cite{itapdb:Ewald1921} as the most famous. In a recent study
for silica,\cite{Brommer2010} linear scaling in the number of particles could be achieved by
using the Wolf summation method\cite{itapdb:Wolf1999} without 
significant loss of accuracy.
By optimizing the damping and truncation of the long-ranged potential
while maintaining energy conservation, simulations can be performed at 
a comparatively small force-field cutoff radius of 8 \AA.

The program \emph{potfit}\cite{potfit, potfit2} generates an effective atomic
interaction force field with the force matching
method\cite{itapdb:Ercolessi1994}, i.e., by adjusting the potential parameters to
optimally reproduce a set of reference data computed in
first-principles calculations.  We have extended the program to
electrostatic interactions and implemented the TS model with Wolf
summation in the optimization algorithm. In this way, it is now
possible to generate force fields for arbitrary metal oxides and
use them in MD simulations, notably in the MD
code IMD\cite{IMD, IMD2}, where the TS model and Wolf summation were
already implemented. \cite{Brommer2010}
The entire approach and the implementation in
\emph{potfit} are described in Sec.~\ref{sec:force}.

This potential generation method was applied to reparametrize
the original TS force field for liquid silica (SiO$_2$).
Because of its technological significance and its high abundance in 
nature,\cite{itapdb:Iler1979} silica has been thoroughly investigated
experimentally\cite{Gaetani1998,Mei2007},
by using \emph{ab initio} calculations\cite{itapdb:Karki2007,Vuilleumier2009,Pasquarello1998} and in
simulations with empirical interaction
potentials\cite{Vashishta1990,itapdb:Beest1990,Horbach1999,itapdb:Tangney2002,Kermode2010}. 
Hence silica is an ideal test case for the new potential generation. 
Furthermore, a force field for liquid magnesia is generated.
Magnesia (MgO) also is one of the most important metal oxides. Not only is 
it one of the simpler oxides (making it a frequently studied test system), 
it is also both ubiquitous and of technological importance.
Hence there have been many experimental\cite{Speziale2001,Boiocchi2001,Peckham1966,Hazen1976} and
theoretical\cite{Karki2006,Alfe2005,Oganov2003,Tangney2003,itapdb:Rowley1998} studies.
By generating a force field for magnesia, we expand the combined use of the TS model 
and Wolf summation and demonstrate its power beyond silica.

We determine basic microstructural, thermodynamic and vibrational properties for silica and magnesia in 
comparison to experimental and first principles results and confirm the quality of this new force field approach by error estimations.
The results are shown in Sec.~\ref{sec:silica} and Sec.~\ref{sec:magnesia} respectively.  
In Sec.~\ref{sec:conclusion} we sum up our results and give an outlook.

\section{Force field}
\label{sec:force}

\subsection{Tangney-Scandolo potential model}

The TS potential contains two contributions: a short-range pair
potential of Morse-Stretch (MS) form, and a long-range part, which
describes the electrostatic interactions between charges and induced
dipoles on the oxygen atoms. The MS interaction between an atom of
type $i$ and an atom of type $j$ has the form
\begin{equation}
\label{eq:MS}
  U^{\text{MS}}_{ij} =
  D_{ij}\left[\exp[\gamma_{ij}(1-\frac{r_{ij}}{r^0_{ij}})] -
    2\exp[\frac{\gamma_{ij}}{2}(1-\frac{r_{ij}}{r^0_{ij}})] \right], 
\end{equation}
with $r_{ij}=\left|\bm{r}_{ij}\right|$,
$\bm{r}_{ij}=\bm{r}_{j}-\bm{r}_{i}$ and the model parameters $D_{ij}$,
$\gamma_{ij}$ and $r^0_{ij}$, which have to be optimized.

The dipole moments depend on the local electric field of the
surrounding charges and dipoles. Hence a self-consistent iterative
solution has to be found. In the TS approach, a dipole moment
$\bm{p}_i^{n}$ at position $\bm{r}_i$ in iteration step $n$ consists
of an induced part due to an electric field $\bm{E}(\bm{r}_i)$ and
a short-range part $\bm{p}^{\text{SR}}_i$ due to the short-range interactions
between charges $q_i$ and $q_j$. Following Rowley
\etal\cite{itapdb:Rowley1998}, this contribution is given by
\begin{equation}
  \label{eq:TS1}
\bm{p}^{\text{SR}}_i = \alpha_i \sum\limits_{j \neq i} \frac{q_j
  \bm{r}_{ij}}{r_{ij}^3}f_{ij}(r_{ij}) 
\end{equation}
with
\begin{equation}
  \label{eq:TS2}
f_{ij}(r_{ij}) = c_{ij} \sum\limits_{k=0}^4 \frac{(b_{ij}r_{ij})^k}{k!}e^{-b_{ij}r_{ij}}.
\end{equation}
$f_{ij}(r_{ij})$ was introduced \emph{ad hoc} to account for multipole effects of 
nearest neighbors and is a function of very short range. 
$b_{ij}$ is the reciprocal of the length scale over which the short-range
interaction comes into play, $c_{ij}$ determines amplitude and sign of this
contribution to the induced moment.
Together with the induced part, one obtains
\begin{equation}
  \label{eq:TS3}
  \bm{p}_i^n=\alpha_i\bm{E}(\bm{r}_i;\{\bm{p}_j^{n-1}\}_{j=1,N},
  \{\bm{r}_j\}_{j=1,N}) +  \bm{p}_i^{\text{SR}}, 
\end{equation}
where $\alpha_i$ is the polarizability of atom $i$ and
$\bm{E}(\bm{r}_i)$ the electric field at position $\bm{r}_i$, which is
determined by the dipole moments $\bm{p}_j$ in the previous iteration
step.
Considering the interactions between charges $U^{\text{qq}}$, between dipole moments $U^{\text{pp}}$
and between a charge and a dipole $U^{\text{pq}}$, the total electrostatic contribution is given by
\begin{equation}
\label{eq:Energy1}
U^{\text{EL}} = U^{\text{qq}} + U^{\text{pq}} + U^{\text{pp}},
\end{equation}
and the total interaction is
\begin{equation}
\label{eq:Energy2}
 U^{\text{tot}} = U^{\text{MS}} + U^{\text{EL}}.
\end{equation}

\subsection{Wolf summation}
\label{subsec:Wolf}

The TS potential consists of pairwise interactions only. 
The electrostatic energies of a condensed system are then described by 
functions with $r^{-n}$ dependence, $n\in\{1,2,3\}$. For point charges 
($r^{-1}$) it is common to apply the Ewald method, where the total Coulomb 
energy of a set of $N$ ions,
\begin{equation}
  \label{eq:Wolf1}
  E^{\text{qq}} = \frac{1}{2} \sum\limits_{i=1}^{N}
  \sum^{N}_{\substack{j=1 \\ j \neq i}}\frac{q_{i}q_{j}}{r_{ij}}, 
\end{equation}
is decomposed in two terms $E_{\bm{r}}^{\text{qq}}$ and
$E_{\bm{k}}^{\text{qq}}$ by inserting a unity of the form
$1=\erfc\!\left(\kappa r\right)+\erf\!\left(\kappa r\right)$ with the
error function
\begin{equation}
  \label{eq:Wolf2}
\erf\!\left(\kappa r\right):=\frac{2}{\sqrt{\pi}}\int\limits_0^{\kappa r}\!dt\,e^{-t^2}.
\end{equation}
The splitting parameter $\kappa$ controls the distribution of energy
contributions between the two terms. The short-ranged erfc term 
is summed up directly, while the smooth erf term is
Fourier transformed and evaluated in reciprocal space. This restricts
the technique to periodic systems. However, the main disadvantage is
the scaling of the computational effort with the number of particles
in the simulation box, which increases as
$O(N^{3/2})$\cite{itapdb:Fincham1994}, even for the optimal choice of
$\kappa$.

Wolf \etal\cite{itapdb:Wolf1999} designed a method with linear scaling properties $O(N)$ for Coulomb
interactions. By taking into account the physical properties of the system,
the reciprocal-space term $E_{\bm{k}}^{\text{qq}}$ is disregarded. It 
can be written as
\begin{align}
  \label{eq:Wolf3}
 E_{\bm k}^{\text{qq}} &= \frac{2\pi}{V} \sum_{\substack{\bm k\ne\bm 0\\ |\bm k|<k_{c}}}
      S(\bm{k})\frac{\exp(-\frac{|\bm k|^{2}}{4\kappa^{2}})}{|\bm k|^{2}}, \\
\intertext{where}
 \label{eq:Wolf4}
 S(\bm{k}) &=  \left| \sum_{j} q_{j}
       \exp(i\bm{k}\cdot \bm{r}_{j})\right|^{2}
\end{align}
is the charge structure factor.
$V$ is the volume of the simulation box. The charge structure factor
is the Fourier transform of the charge-charge autocorrelation
function. In liquid systems and largely also in solids, 
the charges screen each other like in a
cold dense plasma. This means that for small magnitudes of the wave vectors
$\bm{k}$, the charge structure factor is also small. In a previous
study, we showed\cite{Brommer2010} that the reciprocal-space term is
indeed negligible compared to the real-space part, 
provided that the splitting parameter $\kappa$ is chosen
small enough. As, however, a smaller $\kappa$ results in a larger
real-space cutoff $r_c$, the latter has to be chosen two or three
times the size of a typical short-range cutoff radius in metals. So 
a compromise has to be found for MD simulations:  
On the one hand $\kappa$ has to be small to allow disregarding 
the reciprocal-space term, on the other hand it should not be too small
to avoid an oversized cutoff radius.

In addition, a continuous and smooth cutoff of the remaining screened
Coulomb potential $\tilde{E}^{\text{qq}}(r_{ij})=q_iq_j\erfc(\kappa r_{ij}) r_{ij}^{-1}$
is adopted at $r_c$ by shifting the potential so that it goes to zero
smoothly in the first two derivatives at $r=r_c$. We use the Wolf
method for charges and its extension to dipolar interactions. For more 
information about the Wolf summation of dipole contributions
and a detailed analysis of the energy conservation in MD simulations,
see Ref. \onlinecite{Brommer2010}.

\subsection{Wolf dipole error estimation}

To confirm the validity of the Wolf summation for the dipole
contributions, it has to be analyzed whether the dipolar
reciprocal-space term can also be neglected. This approximation is
justified in a system with no significant long-range dipole-correlations like
in ferroelectrics. The reciprocal-space term of the total
energy of $N$ dipole moments $\bm{p}_i$ can be written -- analogously to
Eq.~\eqref{eq:Wolf3} -- as
\begin{equation}
\label{eq:Error1}
E^{\text{pp}}_{\bm{k}}= \frac{2\pi
  Ne^2}{V}\sum\limits_{\bm{k}\neq\bm{0}}^{\infty}
\bm{k}^t\te{Q}(\bm{k})\bm{k}\frac{\exp(-\frac{|\bm
    k|^{2}}{4\kappa^{2}})}{|\bm k|^{2}},   
\end{equation}
where $e$ is the elementary charge and $\te{Q}(\bm{k})$ the dipole
structure factor
\begin{equation}
\label{eq:Error2}
\te{Q}(\bm{k}):=\frac{1}{Ne^2}\sum\limits_{i,j}^{N}\,
 \bm{p}_i\otimes\bm{p}_j\,\exp(i\bm{k}\cdot\bm{r}_{ij}),
\end{equation}
with the normalization factor $1/\sqrt{Ne^2}$. In each system of interest, 
the reciprocal-space term has to be negligible compared to the real-space contribution,
which can be checked by evaluating the dipole structure factor.

\subsection{Implementation in \emph{potfit}}

The programm \emph{potfit} generates an effective atomic interaction
force field solely from \emph{ab initio} reference structures.
The potential parameters are optimized by matching the resulting
forces, energies and stresses to according first-principles values
with the force matching method. All
reference structures used in this study were prepared 
with the plane wave code VASP\cite{Kresse1993,Kresse1996}. 
For $N_m$ particles, reference configuration $m$ provides one energy
$e^0_{m}$, six components of the stress tensor $s^0_{m,l}$ ($l =
1,2,..,6$) and $3N_m$ total force cartesian components $f^0_{m,n}$ ($n =
1,2,...,3N_m$) on $N_m$ atoms. The function
\begin{equation}
\label{eq:Impl1}
Z = w_e Z_e + w_s Z_s + Z_f
\end{equation}
is minimized with respect to the charges, the parameters of the MS interaction Eq.~\eqref{eq:MS}
and the dipolar contribution Eqs.~\eqref{eq:TS1}--\eqref{eq:TS3}. Here
\begin{align}
\label{eq:Impl2}
\begin{split}
 Z_e &= 3\sum\limits_{m=1}^{M} N_m(e_{m}-e^0_{m})^2, \\ 
 Z_s &= \frac{1}{2}\sum\limits_{m=1}^{M}
 \sum\limits_{l=1}^{6}N_m(s_{m,l}-s^0_{m,l})^2, \\ 
 Z_f &= \sum\limits_{m=1}^{M}\sum\limits_{n=1}^{3N_m}
 (f_{m,n}-f^0_{m,n})^2, 
\end{split}
\end{align}
and $e_{m}$, $s_{m,l}$ and $f_{m,n}$ are the corresponding values calculated
with the parametrized force field. $w_e$ and $w_s$ are certain weights 
to balance the different amount of available data for each quantity. 
In the following we assume $M$ reference structures that all consist of 
the same number of particles ($N_m = N$), but in
principle, \emph{potfit} can handle different numbers of particles for each
reference structure. The root mean square (rms) errors,
\begin{equation}
\label{eq:Impl3}
\Delta F_e = \sqrt{\frac{Z_e}{3MN}}~~,~~\Delta F_s = \sqrt{\frac{2Z_s}{MN}}~~~\text{and}~~\Delta F_f = \sqrt{\frac{Z_f}{MN}},
\end{equation}
are first indicators of the quality of
the generated force field. Their magnitudes are independent of
weighting factors, number and sizes of reference structures. 
For the minimization of the potential parameters,
a combination of a stochastic simulated annealing
algorithm\cite{itapdb:Corona1987} and a conjugate-gradient-like deterministic
algorithm\cite{itapdb:Powell1965} is used. 

We have extended the repertory of interactions in \emph{potfit} by
implementing a long-range electrostatic pair-force-routine including
the TS force field with Wolf summation. Earlier we had also implemented
it into the MD code IMD\cite{Brommer2010}, so that potential generation and
simulation can now be used successively. \emph{potfit} is parallelized
using the standard Message Passing Interface (MPI)\cite{itapdb:Gropp1999} by
distributing the reference configurations on the processors. The whole
method is not limited to diatomic metal oxides, but can treat any
oxide.

\emph{potfit} now accepts the TS force field parameters as values to be
optimized. The MS potential Eq.~\eqref{eq:MS} is defined by three parameters for each pair of
interaction partners $ij$. Including the charges $q_i$, the
polarizability $\alpha$ of the oxygen atom and the parameters $b_{ij}$
and $c_{ij}$ (Eqs.~\eqref{eq:TS1}--\eqref{eq:TS3}), 
which only differ from zero in the case $i \neq j$,
there are 14 parameters for a binary oxide. The requirement of charge
neutrality reduces the number of free parameters by one. Thus, there
are 13 parameters to optimize.

\section{Application to silica}
\label{sec:silica}

\subsection{Parametrization}

The new potential generation method has first been applied to liquid silica. We
prepared five MD trajectories using the Wolf summed TS
force field without reparametrization\cite{Brommer2010}. These constant-temperature 
runs were performed at five different temperatures between 2000 and 4000 Kelvin. In 
addition, the volume was slightly lowered to prepare different
pressure conditions.
Then we took several snapshots
following 10 ps of equilibration. In this way, we prepared 47 liquid reference structures
with on average 109 atoms, altogether 5123 atoms. The reference structures show a pressure
spectrum from zero up to 15 GPa. They are used as input configurations 
for the first-principles plane wave code 
VASP, \cite{Kresse1993,Kresse1996} where PAW pseudopotentials\cite{itapdb:Kresse1999}
and a generalized gradient approximation (GGA) of the exchange-correlation 
functional were used. 
With the local-density approximation (LDA) exchange-correlation functional, 
the well-known underestimation of the volume yielded clear deviations 
from experimental data. The approach with GGA and the same reference structures, 
however, did not overestimate the 
volume in MD simulations. Using ultrasoft pseudopotentials to generate 
a reference daase had no noticeable influence on the results. 

The weights in \emph{potfit} were chosen to $w_e = 0.1$ and $w_s =
0.5$, which is consistent with comparable optimization
approaches.\cite{Kermode2010,itapdb:Tangney2002} The resulting parameters, 
however, remained rather unaffected by modification of the weights.
By contrast, the splitting parameter $\kappa$ had to be optimized as described 
in Sec.\,\,II B. Setting $\kappa= 0.02\ \text{\AA}^{-1}$, a cutoff
radius of only 8 \AA\ was found to be sufficient, which is small 
compared to other\cite{Kermode2010,Brommer2010,Carre2007} long-range potential approaches and results in an
additional speedup in simulations. For comparison only, the procedure was repeated
with a cutoff radius of 10 \AA. The obtained results were quite similar.
As the computational effort scales with $r_{\text{cut}}^3$, simulations with a 
cutoff radius of 8 \AA\ are about two times faster than the same settings with 
a cutoff radius of 10 \AA\ (see Fig.~\ref{fig:scaling}). It has to be mentioned
that $\kappa= 0.02\ \text{\AA}^{-1}$ describes a relatively weak damping. This identifies the good native 
screening ability of liquid silica.  

The final set of parameters is shown in Table~\ref{tab:para_si}.
The rms errors are $\Delta F_e = 0.1922$,
$\Delta F_s = 0.0341$ and $\Delta F_f = 1.6211$. 

\begin{table}
   \centering
   \begin{tabular}{c c c c c}
\hline
\hline
\oneline{$q_{\text{Si}}$}{$q_{\text{O}}$}{$\alpha_\text{O}$}{$b_{\text{Si}-\text{O}}$}{$c_{\text{Si}-\text{O}}$} 
\oneline{~1.860 032~}{$-$0.930 016}{~0.020 689~}{4.434 517}{~$-$31.525 678~}
\hline
\oneline{$D_{\text{Si}-\text{Si}}$}{}{$D_{\text{Si}-\text{O}}$}{}{$D_{\text{O}-\text{O}}$}
\oneline{0.000 004}{}{0.100 108}{}{0.028 596}
\hline
\oneline{$\gamma_{\text{Si}-\text{Si}}$}{}{$\gamma_{\text{Si}-\text{O}}$}{}{$\gamma_{\text{O}-\text{O}}$}
\oneline{19.841 872}{}{11.598 884}{}{8.808 762}
\hline
\oneline{$r^0_{\text{Si}-\text{Si}}$}{}{$r^0_{\text{Si}-\text{O}}$}{}{$r^0_{\text{O}-\text{O}}$}
\oneline{5.400 713}{}{2.066 695}{}{3.742 815}
\hline
\hline
\end{tabular}
\caption{Force field parameters for silica, given in IMD units set eV,
  \AA\ and amu (hence charges are multiples of the elementary charge).
} 
\label{tab:para_si}
\end{table}

\subsection{Wolf dipole error estimation}

To analyze reciprocal-space contributions we simulated liquid
silica with 4896 atoms at 3100 K and averaged several observables
over the full simulation time of 200 ps.  The total dipole moment is
$p = 3.47\times10^{-28}$\, Cm, which is small compared to a
fully polarized system and thus can be taken as a fluctuation.
Furthermore, we calculated the dipole structure factor and the
resulting reciprocal-space contribution of the total energy, given in
Eq.~\eqref{eq:Error1}. For the selected damping of $\kappa= 0.02 \,
\text{\AA}^{-1}$ we get
\begin{equation}
\label{eq:estimation_si_1}
\frac{1}{N} E^{\text{pp}}_{\bm{k}}  = 0.09 \, \mu{\text eV}, 
\end{equation}
whereas the real-space contribution of the total energy per atom is
\begin{equation}
\label{eq:estimation_si_2}
\frac{1}{N} E^{\text{tot}}_{\bm{r}}  = 7.20 \, {\text eV}. 
\end{equation}
Hence the reciprocal-space term is indeed insignificant and can be neglected. 

\subsection{Results}
\label{subsec:results_si}

\begin{figure}
  \centering
 \includegraphics{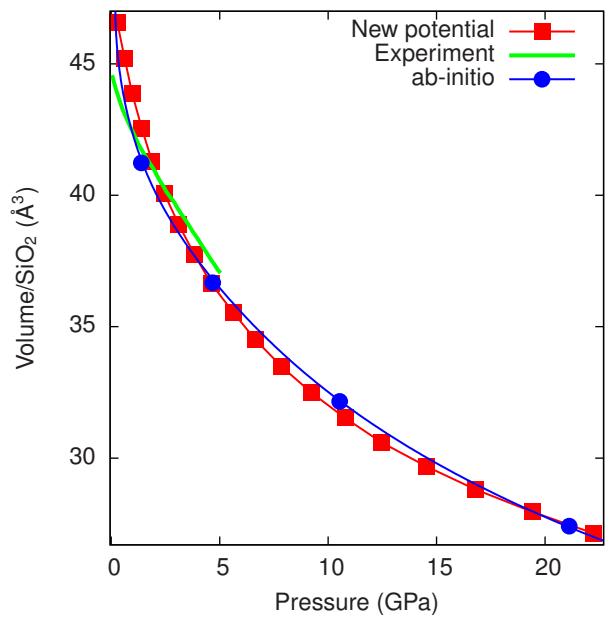}
 \caption{Equation of state of liquid silica at 3100 K for the new potential compared with
   experiment\cite{Gaetani1998} and \emph{ab initio} calculations.\cite{itapdb:Karki2007} }
 \label{fig:eqstate}
\end{figure}

\begin{figure}
  \centering
 \includegraphics{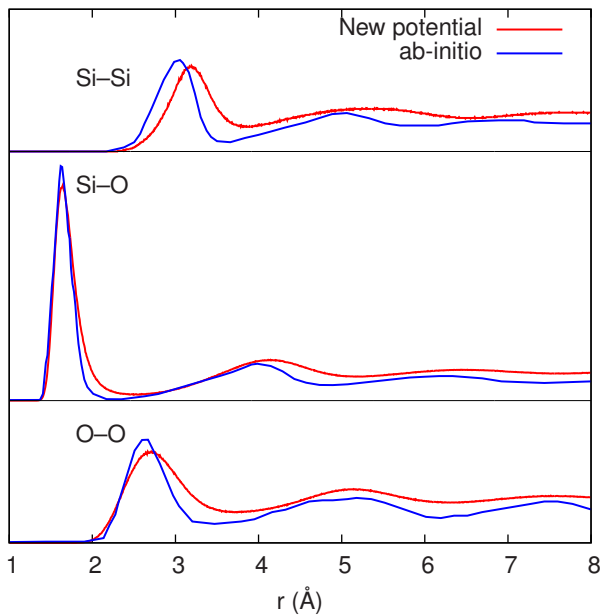}
 \caption{Radial distribution functions for Si--Si, Si--O and O--O at 3000 K compared with 
   \emph{ab initio} calculations.\cite{itapdb:Karki2007} }
 \label{fig:pair}
\end{figure}

\begin{figure}
  \centering
  \includegraphics{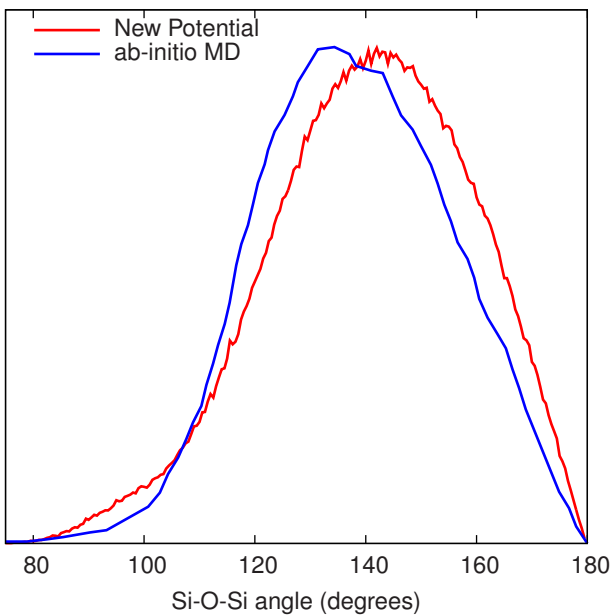}
 \caption{Oxygen centered angle distribution in liquid silica at 2370 K for the new potential compared with an \emph{ab initio}
    MD study. \cite{Vuilleumier2009} }
  \label{fig:angle}
\end{figure}

\begin{figure}
  \centering
  \includegraphics{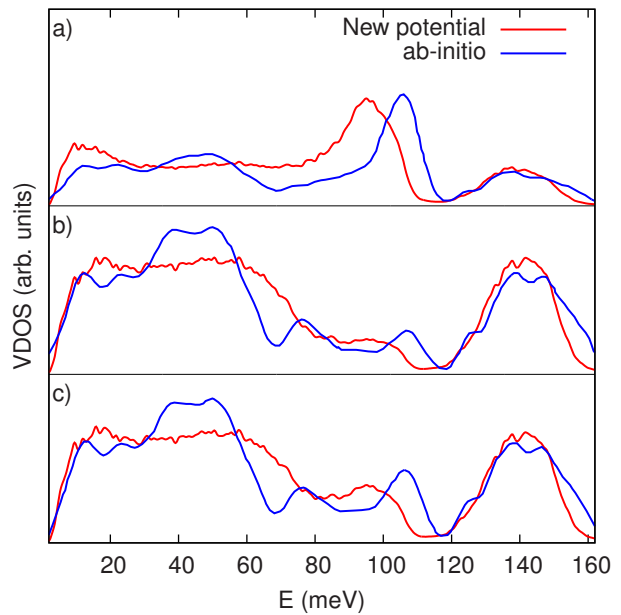}
 \caption{VDOS of amorphous silica at 300 K calculated with the new potential
   compared with an \emph{ab initio} MD study\cite{Pasquarello1998}. a) partial VDOS for silicon atoms,
   b) partial VDOS for oxygen atoms, c) generalized VDOS.}
  \label{fig:vdos}
\end{figure}

We validated the potential by determining thermodynamic, microstructural and vibrational
properties of silica. The goal is to cover different temperature and pressure  
scenarios in order to show the transferability of the new potential. 
The following simulations were all performed with the same initial configuration 
consisting of 4896 atoms (1632 Si and 3264 O).

In Fig.~\ref{fig:eqstate}, we show the equation of state of liquid
silica at 3100 K calculated with the new potential.
Pressures were obtained as averages along constant-volume MD runs 
of 10 ps following 10 ps of equilibration. For comparison, 
emperimental data\cite{Gaetani1998} and first principles
results using VASP\cite{itapdb:Karki2007} are illustrated. 
The equation of state calculated with the new potential coincides
with \emph{ab initio} and experimental results. 

The radial distribution functions for Si--Si, Si--O and O--O at 3000 K with
volume/SiO$_2$ $V_{\text{SiO}_2}=45.80$ \AA$^3$ are evaluated for 100
snapshots taken out of an 100 ps MD run. The averaged curves are given in
Fig.~\ref{fig:pair}. Results calculated with the new potential are in accurate
agreement with \emph{ab initio} data\cite{itapdb:Karki2007}. The Si--Si and O--O curves 
are slightly shifted to larger distances. 

The Si--O--Si angle distribution was determined from 300 MD runs
at 2370 K and zero pressure. The averaged result is
compared with an \emph{ab initio} MD study\cite{Vuilleumier2009} and is depicted in
Fig.~\ref{fig:angle}. As discussed in Ref.~\onlinecite{Vuilleumier2009}, \emph{ab initio} MD
tends to shift the angle distribution of liquid silica to slightly smaller angles with respect to the distribution
generated by empirical MD simulations. 

The new force field was also applied to simulations of amorphous silica, although
its parameters were only optimized with liquid reference structures between 2000 and 4000~K.
For this purpose, the liquid initial structure was cooled down to 300 K at an annealing rate of 0.01 K/fs, which is
recommended by Vollmayr \etal\cite{Vollmayr1996} and also used in Ref. \onlinecite{Kermode2010}. Using different 
annealing rates, however, had no significant impact on the results. The \emph{partial}
vibrational density of states (VDOS) G$_{\text{Si}}$(E) (G$_{\text{O}}$(E)) for silicon (oxygen) 
was obtained by computing the 
Fourier transform of the time-dependent velocity-velocity autocorrelation function from a 100 ps MD trajectory
with the software package nMoldyn\cite{Rog2003}. The \emph{generalized} VDOS G(E) is then calculated by
\begin{equation}
G(E) = \sum\limits_{\mu = \text{Si,O}}^{} \frac{\sigma_{\mu}}{m_{\mu}} G_{\mu}(E)
\end{equation}
with the scattering cross section $\sigma_{\mu}$ and atomic mass $m_{\mu}$ of atom $\mu$.
Fig.~\ref{fig:vdos} shows the partial and generalized VDOS compared with an \emph{ab initio} study from Ref. \onlinecite{Pasquarello1998}. 
The curves were adjusted in order to display the three main bands of
each curve at similar frequencies. To achieve this, 
a global, constant relative frequency scaling for the MD curves,
\begin{equation}
\omega = (1 + \gamma)\omega'
\end{equation}
was introduced, where $\omega'$ is the original eigenfrequency, $\omega$ the scaled frequency, and $\gamma$
the constant relative frequency scaling. For silica we find the optimal $\gamma = 0.3$.
The new potential reproduces the key features of the partial VDOS. There are, however, several smaller deviations from \emph{ab initio}
results: The partial VDOS for silicon overestimates the low-energy band as the main-peak is shifted by around 12 meV to lower energies.
The partial VDOS for ogyxen does not reproduce the band between 35--55 meV and the two smaller peaks at around 77 meV and 108 meV.
These characteristics are also reflected in the generalized VDOS. The relatively large scaling of the MD curves might also be a 
weakness of the new potential. In summary, the new potential is able to qualitatively reflect the lattice 
dynamics, although it was not optimized for amorphous states at 300 K,
but there are limits to its accuracy.  

As a final test, we used the force field for MD studies of $\alpha$-quartz, one of the most important low-pressure crystal structures of SiO$_2$. 
This is a very hard test for the transferability of the potential, considering liquid structures between 2000 and 4000 K were used as reference data. 
In Table~\ref{tab:quartz}, we compare the lattice parameters, Si--O bond length, and the Si--O--Si angle to analytical values\cite{Gibbs2009, Gibbs2006}. 
The force field overestimates the equilibrium volume; the lattice parameters are too large by on average 2.8\%. 
In an MD simulation at 300 K over 10 ns, our force field stabilized the alpha-quartz structure. 
This shows that the new potential does yield reasonable results even under conditions it was not optimized for.

\begin{table}
   \centering
   \begin{tabular}{l l l l l}
\hline
\hline
& \multicolumn{1}{l}{$a$ (\AA)} & \multicolumn{1}{l}{$c$ (\AA)} & \multicolumn{1}{l}{Si--O} (\AA) & \multicolumn{1}{l}{Si--O--Si} ($^{\circ}$) \\
 \hline
\oneline{New potential~~}{5.15}{5.50}{~1.65}{~148.5}
\oneline{Theory}{4.97\cite{Gibbs2009}}{5.39\cite{Gibbs2009}}{~1.61\cite{Gibbs2006}}{~145\cite{Gibbs2006}}
\hline
\hline

  \end{tabular}
\caption{Lattice parameters, Si--O bond length and Si--O--Si angle of $\alpha$-quartz compared with theoretical studies.} 
\label{tab:quartz}
\end{table}

\begin{figure}
  \centering
  \includegraphics{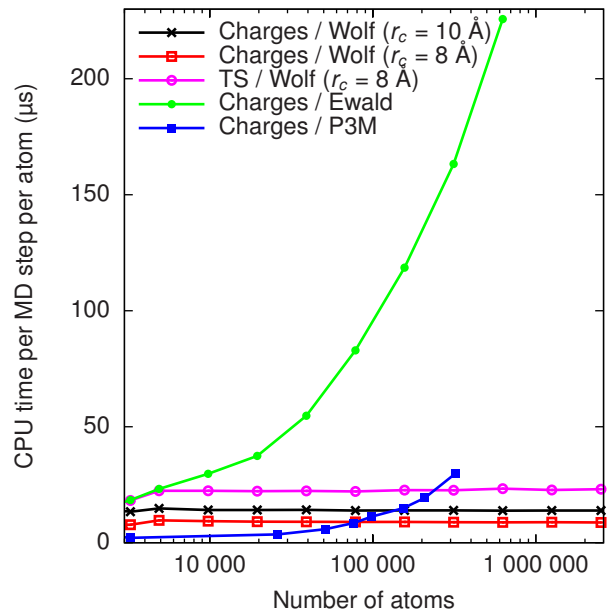}
  \caption{Scaling of computational effort with system size for various long-range interaction techniques. For pure point charges, P3M is fastest for systems
with less than 80 000 atoms. In larger systems, Wolf summation performs better. The TS model also scales linear with system size using Wolf summation 
(it is not implemented in Ewald summation or the P3M method); it is slower by a factor of about 2.6 compared to point charges.}
 \label{fig:scaling}
\end{figure}

We simulated systems with up to 2.5 million atoms in order to probe the linear scaling properties of the model in IMD.
To judge the performance of the Wolf summation, 
the CPU cost of simulating a system of point charges was compared to the P3M\cite{P3M} (Particle-particle/particle-mesh) method, 
as implemented in the MD code ESPResSo\cite{Espresso}. In P3M, the charges associated with the smooth erf term (see Sec.~\ref{subsec:Wolf})
are transferred to a lattice, where then the Poisson equation is solved. 
The CPU time per time step and atom is shown in Fig.~\ref{fig:scaling}. It can be seen clearly that the computational effort with Wolf summation scales perfectly linear 
in our implementation. For smaller systems (fewer than about 80 000 particles), ESPResSo is faster, but as the systems become larger, the $O(N\log N)$ scaling of P3M 
loses to the linear scaling of Wolf. As a reference, we also show the CPU cost of the Ewald method in IMD. All calculations were performed on a single 2.83 GHz Intel Nehalem CPU. 
The computational cost per atom of the TS model is independent of system size; compared to coulombic charges, the TS model in Wolf summation is slower by a factor of 2.6. 
This shows that in general, the number of steps in the self-consistency loop is independent of system size.

\section{Application to magnesia}
\label{sec:magnesia}

\subsection{Parametrization and error estimation}

\begin{table}
   \centering
   \begin{tabular}{c c c c c}
\hline
\hline
\oneline{$q_{\text{Mg}}$}{$q_{\text{O}}$}{$\alpha_\text{O}$}{$b_{\text{Mg}-\text{O}}$}{$c_{\text{Mg}-\text{O}}$} 
\oneline{~1.230 958~}{$-$1.230 958}{~0.045 542~}{3.437 254}{~$-$24.256 585~}
\hline
\oneline{$D_{\text{Mg}-\text{Mg}}$}{}{$D_{\text{Mg}-\text{O}}$}{}{$D_{\text{O}-\text{O}}$}
\oneline{0.000 003}{}{0.100 093}{}{0.000 003}
\hline
\oneline{$\gamma_{\text{Mg}-\text{Mg}}$}{}{$\gamma_{\text{Mg}-\text{O}}$}{}{$\gamma_{\text{O}-\text{O}}$}
\oneline{18.736 878}{}{10.340 058}{}{18.696 021}
\hline
\oneline{$r^0_{\text{Mg}-\text{Mg}}$}{}{$r^0_{\text{Mg}-\text{O}}$}{}{$r^0_{\text{O}-\text{O}}$}
\oneline{6.161 436}{}{2.458 717}{}{6.630 618}
\hline
\hline
\end{tabular}
\caption{Force field parameters for magnesia, given in IMD units set eV,
  \AA\ and amu.
} 
\label{tab:para_mg}
\end{table}

To generate the reference structure database for liquid magnesia, we used snapshots from an MD trajectory with an ad-hoc potential. 
We prepared 80 liquid reference structures with on average 242 atoms, in total 19 360 atoms.
The reference forces, stresses and energies were computed with VASP using PAW pseudopotentials and in the GGA approximation. 
From these, we obtained an intermediate TS potential, which then was again used to generate a new MD trajectory and then a new set of reference structures. 
This procedure was iterated until there was no further significant change in the potential parameters. The pressure spectrum of the final reference structure database
is between 0 and 15 GPa, while the temperature varies between 2000 and 5000 K, to account for the higher melting point of magnesia.

The weights in \emph{potfit} again were set to $w_e = 0.1$ and $w_s =
0.5$. A cutoff radius of 8 \AA\ was sufficient when choosing
$\kappa= 0.1\ \text{\AA}^{-1}$. The final set of parameters is shown in Table~\ref{tab:para_mg}.
The rms errors are $\Delta F_e = 0.1170$, $\Delta F_s = 0.0228$ and $\Delta F_f = 0.5295$.

The following values for the Wolf dipole error estimation again were obtained by averaging over a simulation time of 200 ps.
In liquid magnesia with 5832 atoms at 5000 K, the total dipole moment is $p = 4.76\times 10^{-33}$\, Cm,
which is even smaller than that of silica. Also the reciprocal-space term is again really insignificant. For 
$\kappa= 0.1 \,\text{\AA}^{-1}$ we get
\begin{equation}
\label{eq:estimation_mg_1}
\frac{1}{N} E^{\text{pp}}_{\bm{k}}  = 0.15 \, \mu{\text eV}, 
\end{equation}
whereas the real-space contribution of the total energy per atom is
\begin{equation}
\label{eq:estimation_mg_2}
\frac{1}{N} E^{\text{tot}}_{\bm{r}}  = 6.30 \, {\text eV}. 
\end{equation} 

\subsection{Results}
\label{subsec:results_mg}

\begin{figure}
  \centering
 \includegraphics{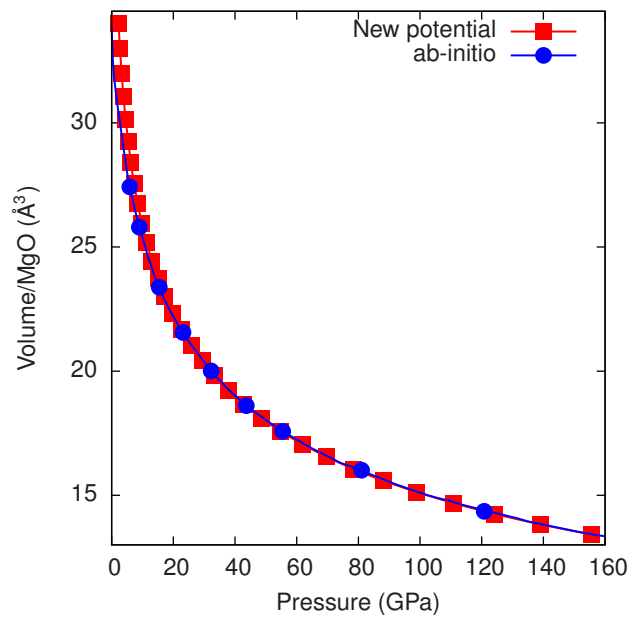}
 \caption{Equation of state of liquid magnesia at 5000 K for the new potential compared with
   \emph{ab initio} calculations.\cite{Karki2006} }
 \label{fig:MgO_eqstate}
\end{figure}

\begin{figure}
  \centering
 \includegraphics{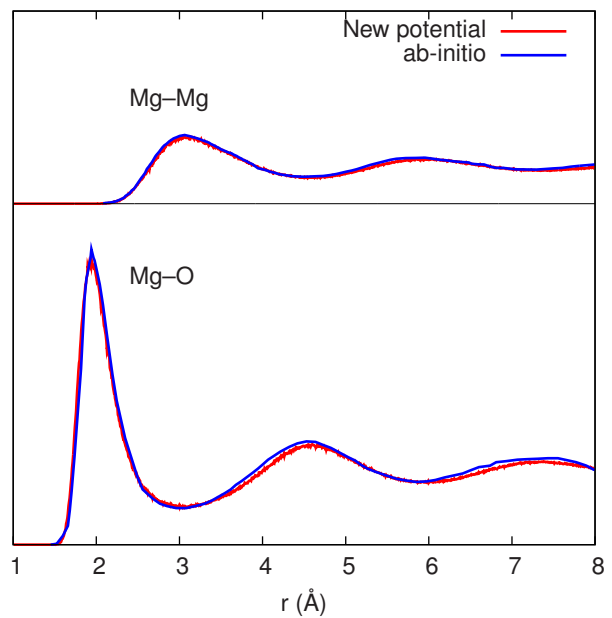}
 \caption{Radial distribution functions for Mg--Mg (which is very similar to O--O case) and Mg--O at 3000 K, 
   compared with \emph{ab initio} calculations.\cite{Karki2006} }
 \label{fig:MgO_pair}
\end{figure}

\begin{figure}
  \centering
  \includegraphics{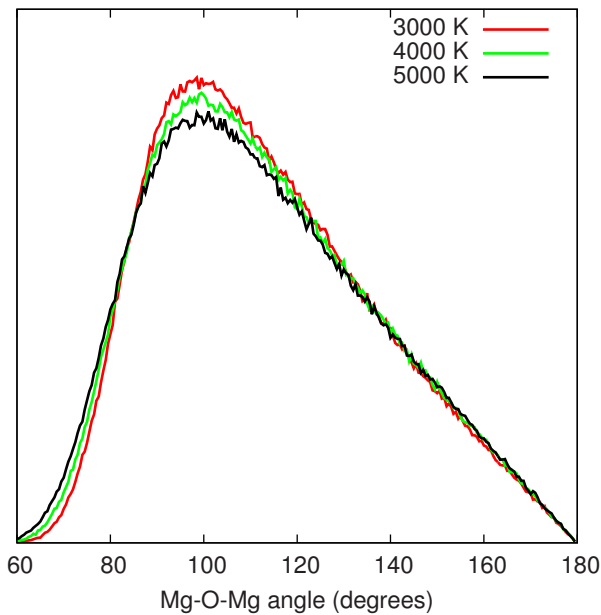}
 \caption{Oxygen centered angle distribution in liquid magnesia for the new potential for 3000 K, 4000 K and 5000 K.}
  \label{fig:MgO_angle}
\end{figure}

\begin{figure}
  \centering
  \includegraphics{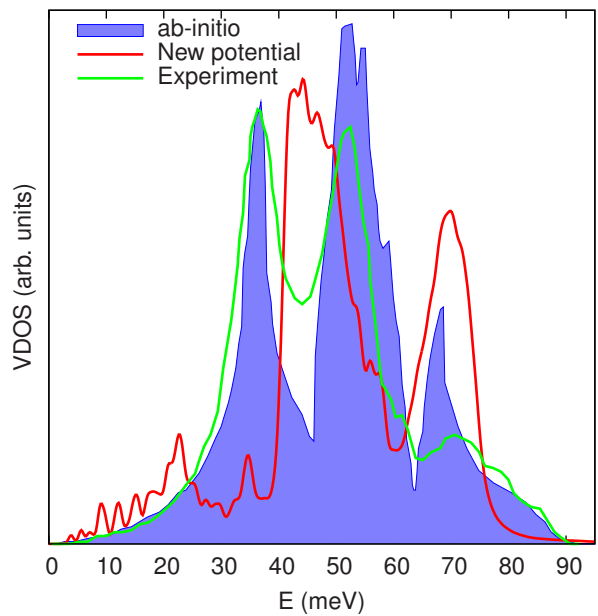}
 \caption{Generalized VDOS of periclase at 300 K calculated with the new potential
   compared with \emph{ab initio} calculations\cite{Ghose2006} and an experimental study.\cite{Bosak2005} }
  \label{fig:MgO_vdos}
\end{figure}

The new potential is validated by determining thermodynamic and microstructural properties of liquid magnesia. 
We also investigated the transferability of the new potential beyond the optimized temperature range by modelling
the most important crystal structure periclase (NaCl-type). The following simulations were all performed with the same 
initial configuration consisting of 5832 atoms (2916 Mg and 2916 O).

The equation of state, obtained as in silica, is shown in Fig.~\ref{fig:MgO_eqstate}. Due to the 
high melting point of magnesia we chose a temperature of 5000 K. The result is compared with a first-principles study using
VASP\cite{Karki2006}. Although the new potential is optimized with a reference database
having a pressure spectrum from zero to 15 GPa, it is able to reproduce the ab initio results quite accurate up to 160 GPa. 

In Fig.~\ref{fig:MgO_pair} the radial distribution function for Mg--Mg (which is very similar to the O--O function) and Mg--O is 
depicted at 3000 K with volume/MgO $V_{\text{MgO}}=27.76\ \text{\AA}^3$. Our results coincide precisely with 
\emph{ab initio} data\cite{Karki2006}. 

The Mg--O--Mg angle distribution was determined from several MD runs at $V_{\text{MgO}}=33.99\ \text{\AA}^3$ and
three different temperatures 3000 K, 4000 K and 5000 K. The curves look similar, they show a maximum at 100$^{\circ}$. 
So the interatomic angles in the magnesia melt are about 10$^{\circ}$ greater than in periclase. Both distribution
studies follow from an averaging in the same way as for silica. 

Finally we applied the new force field to simulations of periclase at 300 K, although its parameters were only 
optimized with liquid reference structures between 2000 and 5000 K. The lattice constant is in good agreement 
with recent \emph{ab-inito} and experimental studies, as presented in Table~\ref{tab:periklas}. 

In the same
way as for silica, we obtained the generalized VDOS of periclase. Fig.~\ref{fig:MgO_vdos} shows a comparison with
\emph{ab initio} calculations\cite{Ghose2006} and an experimental\cite{Bosak2005} study. A frequency scaling of
only $\gamma = 0.03$ is applied. The new potential is able to qualitatively reproduce the key features, but there a
two weak points: Firstly, the main peak at around 53 meV is shifted to lower frequencies by around 10 meV. 
Secondly, the peak at 36 meV -- originating from the partial VDOS for oxygen -- is only weakly reproduced. 
In summary, the new force can give only qualitative results in this temperature range, for which it was not 
optimized.  

Not surprisingly, the scaling properties of simulations of magnesia are the same as those of silica: The computational effort scales 
linear in the number of particles. Also, the TS model is slower than constant charges by a factor independent of system size.

\begin{table}
   \centering
   \begin{tabular}{l d}
\hline
\hline
\twoline{New Potential}{4.214}
\hline
\twoline{Exp.\cite{Speziale2001}}{4.212}
\hline
\twoline{Exp.\cite{Boiocchi2001}}{4.211}
\hline
\twoline{ab initio (GGA)\cite{Alfe2005}}{4.234}
\hline
\twoline{ab initio (LDA)\cite{Oganov2003}}{4.240}
\hline
\hline
\end{tabular}
\caption{Lattice constant $a$ [\AA] of periclase determined with the new potential compared to \emph{ab initio} and experimental studies.} 
\label{tab:periklas}
\end{table}\section{Conclusion}
\label{sec:conclusion}

In this work, we presented the extension of the program \emph{potfit} to 
electrostatic interactions. It can be used to generate a force field for any condensed oxygen system
by force matching to \emph{ab initio} reference data. The system of
interest only has to have an electrostatic screening disposition to apply Wolf summation. 
We have demonstrated that the approach using the TS potential model yields accurate results for liquid silica and magnesia.
Probing solid phases, which were not included in the reference database during the force field optimization, shows
that qualitative studies are possible; quantitative analysis, however, should be done carefully.

A new force field for crystalline $\alpha$-alumina is currently in preparation\cite{Hocker2011} using the methods
introduced in the present publication. In this process many crystalline structures are included in the reference database
for optimizing the potential parameters. This yields a highly accurate description of crystalline properties.

The TS potential is a pure pair term potential. Hence simulations are less
expensive compard to other approaches with many-body potentials like
a three-body interaction approach. Nevertheless the many-body character of the electrostatics is
inclued in the self-consistent dipole iteration approach. 
Due to the Wolf summation, we achieve linear scaling in the 
number of particles, which makes it possible to investigate larger system sizes and timescales. 

Generating a force field for liquid silica is an ideal test case for the new approach, because there
are several recent theoretical studies\cite{itapdb:Tangney2002,Kermode2010,Brommer2010} using the
TS potential model. Beyond silica, we have presented a new force field for liquid magnesia, which yields generally
accurate results. Our work shows that a crude starting potential suffices to determine
high-quality potentials, which makes the new
force field approach applicable for other systems of interest. 

In comparison to former simulation approaches with long-range interactions, we 
present force fields with miminal cutoff radius. This results in faster simulations.   
The short cutoff shows that liquid silica and magnesia can be
adequately described by the Wolf approximation.

We hope that our new force field generation approach enhances the investigation of condensed
oxide systems, which could not yet be adequately investigated with MD simulations, due to 
large system sizes or boundary condition restrictions. 
\begin{acknowledgments}
  The authors thank Daniel Schopf, Franz Gähler, Holger Euchner and Andreas Chatzopoulos for many helpful discussions. 
  We are indebted to Florian Fahrenberger for the collaboration concerning the ESPResSo MD code. 
  Support from the DFG through Collaborative Research Centre 716, Project B.1 is gratefully acknowledged.
\end{acknowledgments}


\begin{thebibliography}{46}%
\makeatletter
\providecommand \@ifxundefined [1]{%
 \@ifx{#1\undefined}
}%
\providecommand \@ifnum [1]{%
 \ifnum #1\expandafter \@firstoftwo
 \else \expandafter \@secondoftwo
 \fi
}%
\providecommand \@ifx [1]{%
 \ifx #1\expandafter \@firstoftwo
 \else \expandafter \@secondoftwo
 \fi
}%
\providecommand \natexlab [1]{#1}%
\providecommand \enquote  [1]{``#1''}%
\providecommand \bibnamefont  [1]{#1}%
\providecommand \bibfnamefont [1]{#1}%
\providecommand \citenamefont [1]{#1}%
\providecommand \href@noop [0]{\@secondoftwo}%
\providecommand \href [0]{\begingroup \@sanitize@url \@href}%
\providecommand \@href[1]{\@@startlink{#1}\@@href}%
\providecommand \@@href[1]{\endgroup#1\@@endlink}%
\providecommand \@sanitize@url [0]{\catcode `\\12\catcode `\$12\catcode
  `\&12\catcode `\#12\catcode `\^12\catcode `\_12\catcode `\%12\relax}%
\providecommand \@@startlink[1]{}%
\providecommand \@@endlink[0]{}%
\providecommand \url  [0]{\begingroup\@sanitize@url \@url }%
\providecommand \@url [1]{\endgroup\@href {#1}{\urlprefix }}%
\providecommand \urlprefix  [0]{URL }%
\providecommand \Eprint [0]{\href }%
\providecommand \doibase [0]{http://dx.doi.org/}%
\providecommand \selectlanguage [0]{\@gobble}%
\providecommand \bibinfo  [0]{\@secondoftwo}%
\providecommand \bibfield  [0]{\@secondoftwo}%
\providecommand \translation [1]{[#1]}%
\providecommand \BibitemOpen [0]{}%
\providecommand \bibitemStop [0]{}%
\providecommand \bibitemNoStop [0]{.\EOS\space}%
\providecommand \EOS [0]{\spacefactor3000\relax}%
\providecommand \BibitemShut  [1]{\csname bibitem#1\endcsname}%
\let\auto@bib@innerbib\@empty
%
\bibitem [{\citenamefont {{Mei}}, \citenamefont {{Benmore}},\ and\
  \citenamefont {{Weber}}(2007)}]{Mei2007}%
  \BibitemOpen
  \bibfield  {author} {\bibinfo {author} {\bibfnamefont {Q.}~\bibnamefont
  {{Mei}}}, \bibinfo {author} {\bibfnamefont {C.~J.}\ \bibnamefont
  {{Benmore}}}, \ and\ \bibinfo {author} {\bibfnamefont {J.~K.~R.}\
  \bibnamefont {{Weber}}},\ }\href@noop {} {\bibfield  {journal} {\bibinfo
  {journal} {Phys.\ Rev.\ Lett.}\ }\textbf {\bibinfo {volume} {98}},\ \bibinfo
  {pages} {057802} (\bibinfo {year} {2007})}\BibitemShut {NoStop}%
\bibitem [{\citenamefont {{Speziale}}, \citenamefont {{Zha}},\ and\
  \citenamefont {{Duffy}}(2001)}]{Speziale2001}%
  \BibitemOpen
  \bibfield  {author} {\bibinfo {author} {\bibfnamefont {S.}~\bibnamefont
  {{Speziale}}}, \bibinfo {author} {\bibfnamefont {C.-S.}\ \bibnamefont
  {{Zha}}}, \ and\ \bibinfo {author} {\bibfnamefont {T.~S.}\ \bibnamefont
  {{Duffy}}},\ }\href@noop {} {\bibfield  {journal} {\bibinfo  {journal}
  {J.~Geophys.\ Res.}\ }\textbf {\bibinfo {volume} {106}},\ \bibinfo {pages}
  {515} (\bibinfo {year} {2001})}\BibitemShut {NoStop}%
\bibitem [{\citenamefont {{Boiocchi}}\ \emph {et~al.}(2001)\citenamefont
  {{Boiocchi}}, \citenamefont {{Caucia}}, \citenamefont {{Merli}},
  \citenamefont {{Prella}},\ and\ \citenamefont {{Ungaretti}}}]{Boiocchi2001}%
  \BibitemOpen
  \bibfield  {author} {\bibinfo {author} {\bibfnamefont {M.}~\bibnamefont
  {{Boiocchi}}}, \bibinfo {author} {\bibfnamefont {F.}~\bibnamefont
  {{Caucia}}}, \bibinfo {author} {\bibfnamefont {M.}~\bibnamefont {{Merli}}},
  \bibinfo {author} {\bibfnamefont {D.}~\bibnamefont {{Prella}}}, \ and\
  \bibinfo {author} {\bibfnamefont {L.}~\bibnamefont {{Ungaretti}}},\
  }\href@noop {} {\bibfield  {journal} {\bibinfo  {journal} {Eur. J. Mineral.}\
  }\textbf {\bibinfo {volume} {13}},\ \bibinfo {pages} {871} (\bibinfo {year}
  {2001})}\BibitemShut {NoStop}%
\bibitem [{\citenamefont {{Karki}}, \citenamefont {{Bhattarai}},\ and\
  \citenamefont {{Stixrude}}(2007)}]{itapdb:Karki2007}%
  \BibitemOpen
  \bibfield  {author} {\bibinfo {author} {\bibfnamefont {B.~B.}\ \bibnamefont
  {{Karki}}}, \bibinfo {author} {\bibfnamefont {D.}~\bibnamefont
  {{Bhattarai}}}, \ and\ \bibinfo {author} {\bibfnamefont {L.}~\bibnamefont
  {{Stixrude}}},\ }\href@noop {} {\bibfield  {journal} {\bibinfo  {journal}
  {Phys.\ Rev.~B}\ }\textbf {\bibinfo {volume} {76}},\ \bibinfo {pages}
  {104205} (\bibinfo {year} {2007})}\BibitemShut {NoStop}%
\bibitem [{\citenamefont {{Vuilleumier}}, \citenamefont {{Sator}},\ and\
  \citenamefont {{Guillot}}(2009)}]{Vuilleumier2009}%
  \BibitemOpen
  \bibfield  {author} {\bibinfo {author} {\bibfnamefont {R.}~\bibnamefont
  {{Vuilleumier}}}, \bibinfo {author} {\bibfnamefont {N.}~\bibnamefont
  {{Sator}}}, \ and\ \bibinfo {author} {\bibfnamefont {B.}~\bibnamefont
  {{Guillot}}},\ }\href@noop {} {\bibfield  {journal} {\bibinfo  {journal}
  {Geochim.\ Cosmochim.\ Acta}\ }\textbf {\bibinfo {volume} {73}},\ \bibinfo
  {pages} {6313} (\bibinfo {year} {2009})}\BibitemShut {NoStop}%
\bibitem [{\citenamefont {{Kermode}}\ \emph {et~al.}(2010)\citenamefont
  {{Kermode}}, \citenamefont {{Cereda}}, \citenamefont {{Tangney}},\ and\
  \citenamefont {{De Vita}}}]{Kermode2010}%
  \BibitemOpen
  \bibfield  {author} {\bibinfo {author} {\bibfnamefont {J.~R.}\ \bibnamefont
  {{Kermode}}}, \bibinfo {author} {\bibfnamefont {S.}~\bibnamefont {{Cereda}}},
  \bibinfo {author} {\bibfnamefont {P.}~\bibnamefont {{Tangney}}}, \ and\
  \bibinfo {author} {\bibfnamefont {A.}~\bibnamefont {{De Vita}}},\ }\href@noop
  {} {\bibfield  {journal} {\bibinfo  {journal} {J.~Chem.\ Phys.}\ }\textbf
  {\bibinfo {volume} {133}},\ \bibinfo {pages} {094102} (\bibinfo {year}
  {2010})}\BibitemShut {NoStop}%
\bibitem [{\citenamefont {{Karki}}, \citenamefont {{Bhattarai}},\ and\
  \citenamefont {{Stixrude}}(2006)}]{Karki2006}%
  \BibitemOpen
  \bibfield  {author} {\bibinfo {author} {\bibfnamefont {B.~B.}\ \bibnamefont
  {{Karki}}}, \bibinfo {author} {\bibfnamefont {D.}~\bibnamefont
  {{Bhattarai}}}, \ and\ \bibinfo {author} {\bibfnamefont {L.}~\bibnamefont
  {{Stixrude}}},\ }\href@noop {} {\bibfield  {journal} {\bibinfo  {journal}
  {Phys.\ Rev.~B}\ }\textbf {\bibinfo {volume} {73}},\ \bibinfo {pages}
  {174208} (\bibinfo {year} {2006})}\BibitemShut {NoStop}%
\bibitem [{\citenamefont {{Alfe}}(2005)}]{Alfe2005}%
  \BibitemOpen
  \bibfield  {author} {\bibinfo {author} {\bibfnamefont {D.}~\bibnamefont
  {{Alfe}}},\ }\href@noop {} {\bibfield  {journal} {\bibinfo  {journal} {Phys.\
  Rev.\ Lett.}\ }\textbf {\bibinfo {volume} {94}},\ \bibinfo {pages} {235701}
  (\bibinfo {year} {2005})}\BibitemShut {NoStop}%
\bibitem [{\citenamefont {{Tangney}}\ and\ \citenamefont
  {{Scandolo}}(2002)}]{itapdb:Tangney2002}%
  \BibitemOpen
  \bibfield  {author} {\bibinfo {author} {\bibfnamefont {P.}~\bibnamefont
  {{Tangney}}}\ and\ \bibinfo {author} {\bibfnamefont {S.}~\bibnamefont
  {{Scandolo}}},\ }\href@noop {} {\bibfield  {journal} {\bibinfo  {journal}
  {J.~Chem.\ Phys.}\ }\textbf {\bibinfo {volume} {117}},\ \bibinfo {pages}
  {8898} (\bibinfo {year} {2002})}\BibitemShut {NoStop}%
\bibitem [{\citenamefont {{Herzbach}}, \citenamefont {{Binder}},\ and\
  \citenamefont {{Muser}}(2005)}]{itapdb:Herzbach2005}%
  \BibitemOpen
  \bibfield  {author} {\bibinfo {author} {\bibfnamefont {D.}~\bibnamefont
  {{Herzbach}}}, \bibinfo {author} {\bibfnamefont {K.}~\bibnamefont
  {{Binder}}}, \ and\ \bibinfo {author} {\bibfnamefont {M.~H.}\ \bibnamefont
  {{Muser}}},\ }\href@noop {} {\bibfield  {journal} {\bibinfo  {journal}
  {J.~Chem.\ Phys.}\ }\textbf {\bibinfo {volume} {123}},\ \bibinfo {pages}
  {124711} (\bibinfo {year} {2005})}\BibitemShut {NoStop}%
\bibitem [{\citenamefont {{Ewald}}(1921)}]{itapdb:Ewald1921}%
  \BibitemOpen
  \bibfield  {author} {\bibinfo {author} {\bibfnamefont {P.~P.}\ \bibnamefont
  {{Ewald}}},\ }\href@noop {} {\bibfield  {journal} {\bibinfo  {journal} {Ann.\
  Phys.\ (Leipzig)}\ }\textbf {\bibinfo {volume} {369}},\ \bibinfo {pages} {253}
  (\bibinfo {year} {1921})}\BibitemShut {NoStop}%
\bibitem [{\citenamefont {{Brommer}}\ \emph {et~al.}(2010)\citenamefont
  {{Brommer}}, \citenamefont {{Beck}}, \citenamefont {{Chatzopoulos}},
  \citenamefont {{Gähler}}, \citenamefont {{Roth}},\ and\ \citenamefont
  {{Trebin}}}]{Brommer2010}%
  \BibitemOpen
  \bibfield  {author} {\bibinfo {author} {\bibfnamefont {P.}~\bibnamefont
  {{Brommer}}}, \bibinfo {author} {\bibfnamefont {P.}~\bibnamefont {{Beck}}},
  \bibinfo {author} {\bibfnamefont {A.}~\bibnamefont {{Chatzopoulos}}},
  \bibinfo {author} {\bibfnamefont {F.}~\bibnamefont {{Gähler}}}, \bibinfo
  {author} {\bibfnamefont {J.}~\bibnamefont {{Roth}}}, \ and\ \bibinfo {author}
  {\bibfnamefont {H.-R.}\ \bibnamefont {{Trebin}}},\ }\href@noop {} {\bibfield
  {journal} {\bibinfo  {journal} {J.~Chem.\ Phys.}\ }\textbf {\bibinfo {volume}
  {132}},\ \bibinfo {pages} {194109} (\bibinfo {year} {2010})}\BibitemShut
  {NoStop}%
\bibitem [{\citenamefont {{Wolf}}\ \emph {et~al.}(1999)\citenamefont {{Wolf}},
  \citenamefont {{Keblinski}}, \citenamefont {{Phillpot}},\ and\ \citenamefont
  {{Eggebrecht}}}]{itapdb:Wolf1999}%
  \BibitemOpen
  \bibfield  {author} {\bibinfo {author} {\bibfnamefont {D.}~\bibnamefont
  {{Wolf}}}, \bibinfo {author} {\bibfnamefont {P.}~\bibnamefont {{Keblinski}}},
  \bibinfo {author} {\bibfnamefont {S.~R.}\ \bibnamefont {{Phillpot}}}, \ and\
  \bibinfo {author} {\bibfnamefont {J.}~\bibnamefont {{Eggebrecht}}},\
  }\href@noop {} {\bibfield  {journal} {\bibinfo  {journal} {J.~Chem.\ Phys.}\
  }\textbf {\bibinfo {volume} {110}},\ \bibinfo {pages} {8254} (\bibinfo {year}
  {1999})}\BibitemShut {NoStop}%
\bibitem [{\citenamefont {{Brommer}}\ and\ \citenamefont
  {{G{\"a}hler}}(2007)}]{potfit}%
  \BibitemOpen
  \bibfield  {author} {\bibinfo {author} {\bibfnamefont {P.}~\bibnamefont
  {{Brommer}}}\ and\ \bibinfo {author} {\bibfnamefont {F.}~\bibnamefont
  {{G{\"a}hler}}},\ }\href@noop {} {\bibfield  {journal} {\bibinfo  {journal}
  {Modelling Simul. Mater. Sci. Eng.}\ }\textbf {\bibinfo {volume} {15}},\
  \bibinfo {pages} {295} (\bibinfo {year} {2007})},\ \bibinfo {note} {{\tt
  http://www.itap.physik.uni-stuttgart.de/\~{}imd/potfit/}}\BibitemShut
  {NoStop}%
\bibitem [{\citenamefont {{Brommer}}\ and\ \citenamefont
  {{G{\"a}hler}}(2006)}]{potfit2}%
  \BibitemOpen
  \bibfield  {author} {\bibinfo {author} {\bibfnamefont {P.}~\bibnamefont
  {{Brommer}}}\ and\ \bibinfo {author} {\bibfnamefont {F.}~\bibnamefont
  {{G{\"a}hler}}},\ }\href@noop {} {\bibfield  {journal} {\bibinfo  {journal}
  {Phil.\ Mag.}\ }\textbf {\bibinfo {volume} {86}},\ \bibinfo {pages} {753}
  (\bibinfo {year} {2006})}\BibitemShut {NoStop}%
\bibitem [{\citenamefont {{Ercolessi}}\ and\ \citenamefont
  {{Adams}}(1994)}]{itapdb:Ercolessi1994}%
  \BibitemOpen
  \bibfield  {author} {\bibinfo {author} {\bibfnamefont {F.}~\bibnamefont
  {{Ercolessi}}}\ and\ \bibinfo {author} {\bibfnamefont {J.~B.}\ \bibnamefont
  {{Adams}}},\ }\href@noop {} {\bibfield  {journal} {\bibinfo  {journal}
  {Europhys.\ Lett.}\ }\textbf {\bibinfo {volume} {26}},\ \bibinfo {pages}
  {583} (\bibinfo {year} {1994})}\BibitemShut {NoStop}%
\bibitem [{\citenamefont {{Stadler}}, \citenamefont {{Mikulla}},\ and\
  \citenamefont {{Trebin}}(1997)}]{IMD}%
  \BibitemOpen
  \bibfield  {author} {\bibinfo {author} {\bibfnamefont {J.}~\bibnamefont
  {{Stadler}}}, \bibinfo {author} {\bibfnamefont {R.}~\bibnamefont
  {{Mikulla}}}, \ and\ \bibinfo {author} {\bibfnamefont {H.-R.}\ \bibnamefont
  {{Trebin}}},\ }\href@noop {} {\bibfield  {journal} {\bibinfo  {journal}
  {Int.\ J.\ Mod.\ Phys.~C}\ }\textbf {\bibinfo {volume} {8}},\ \bibinfo
  {pages} {1131} (\bibinfo {year} {1997})},\ \bibinfo {note} {{\tt
  http://www.itap.physik.uni-stuttgart.de/\~{}imd/}}\BibitemShut {NoStop}%
\bibitem [{\citenamefont {{Roth}}, \citenamefont {{Gähler}},\ and\
  \citenamefont {{Trebin}}(2000)}]{IMD2}%
  \BibitemOpen
  \bibfield  {author} {\bibinfo {author} {\bibfnamefont {J.}~\bibnamefont
  {{Roth}}}, \bibinfo {author} {\bibfnamefont {F.}~\bibnamefont {{Gähler}}}, \
  and\ \bibinfo {author} {\bibfnamefont {H.-R.}\ \bibnamefont {{Trebin}}},\
  }\href@noop {} {\bibfield  {journal} {\bibinfo  {journal} {Int.\ J.\ Mod.\
  Phys.~C}\ }\textbf {\bibinfo {volume} {11}},\ \bibinfo {pages} {317}
  (\bibinfo {year} {2000})}\BibitemShut {NoStop}%
\bibitem [{\citenamefont {{Iler}}(1979)}]{itapdb:Iler1979}%
  \BibitemOpen
  \bibfield  {author} {\bibinfo {author} {\bibfnamefont {R.~K.}\ \bibnamefont
  {{Iler}}},\ }\href@noop {} {\emph {\bibinfo {title} {The chemistry of
  silica}}},\ A {W}iley-{I}nterscience publication\ (\bibinfo  {publisher}
  {John Wiley \& Sons},\ \bibinfo {year} {1979})\BibitemShut {NoStop}%
\bibitem [{\citenamefont {{Gaetani}}, \citenamefont {{Asimow}},\ and\
  \citenamefont {{Stolper}}(1998)}]{Gaetani1998}%
  \BibitemOpen
  \bibfield  {author} {\bibinfo {author} {\bibfnamefont {G.~A.}\ \bibnamefont
  {{Gaetani}}}, \bibinfo {author} {\bibfnamefont {P.~D.}\ \bibnamefont
  {{Asimow}}}, \ and\ \bibinfo {author} {\bibfnamefont {E.~M.}\ \bibnamefont
  {{Stolper}}},\ }\href@noop {} {\bibfield  {journal} {\bibinfo  {journal}
  {Geochim.\ Cosmochim.\ Acta}\ }\textbf {\bibinfo {volume} {62}},\ \bibinfo
  {pages} {2499} (\bibinfo {year} {1998})}\BibitemShut {NoStop}%
\bibitem [{\citenamefont {{Pasquarello}}, \citenamefont {{Sarnthein}},\ and\
  \citenamefont {{Car}}(1998)}]{Pasquarello1998}%
  \BibitemOpen
  \bibfield  {author} {\bibinfo {author} {\bibfnamefont {A.}~\bibnamefont
  {{Pasquarello}}}, \bibinfo {author} {\bibfnamefont {J.}~\bibnamefont
  {{Sarnthein}}}, \ and\ \bibinfo {author} {\bibfnamefont {R.}~\bibnamefont
  {{Car}}},\ }\href@noop {} {\bibfield  {journal} {\bibinfo  {journal} {Phys.\
  Rev.~B}\ }\textbf {\bibinfo {volume} {57}},\ \bibinfo {pages} {14133}
  (\bibinfo {year} {1998})}\BibitemShut {NoStop}%
\bibitem [{\citenamefont {{Vashishta}}\ \emph {et~al.}(1990)\citenamefont
  {{Vashishta}}, \citenamefont {{Kalia}}, \citenamefont {{Rino}},\ and\
  \citenamefont {{Ebbsjö}}}]{Vashishta1990}%
  \BibitemOpen
  \bibfield  {author} {\bibinfo {author} {\bibfnamefont {P.}~\bibnamefont
  {{Vashishta}}}, \bibinfo {author} {\bibfnamefont {R.}~\bibnamefont
  {{Kalia}}}, \bibinfo {author} {\bibfnamefont {J.}~\bibnamefont {{Rino}}}, \
  and\ \bibinfo {author} {\bibfnamefont {I.}~\bibnamefont {{Ebbsjö}}},\
  }\href@noop {} {\bibfield  {journal} {\bibinfo  {journal} {Phys.\ Rev.~B}\
  }\textbf {\bibinfo {volume} {41}},\ \bibinfo {pages} {12197} (\bibinfo {year}
  {1990})}\BibitemShut {NoStop}%
\bibitem [{\citenamefont {van {Beest}}, \citenamefont {{Kramer}},\ and\
  \citenamefont {van {Santen}}(1990)}]{itapdb:Beest1990}%
  \BibitemOpen
  \bibfield  {author} {\bibinfo {author} {\bibfnamefont {B.~W.~H.}\
  \bibnamefont {van {Beest}}}, \bibinfo {author} {\bibfnamefont {G.~J.}\
  \bibnamefont {{Kramer}}}, \ and\ \bibinfo {author} {\bibfnamefont {R.~A.}\
  \bibnamefont {van {Santen}}},\ }\href@noop {} {\bibfield  {journal} {\bibinfo
   {journal} {Phys.\ Rev.\ Lett.}\ }\textbf {\bibinfo {volume} {64}},\ \bibinfo
  {pages} {1955} (\bibinfo {year} {1990})}\BibitemShut {NoStop}%
\bibitem [{\citenamefont {{Horbach}}\ and\ \citenamefont
  {{Kob}}(1999)}]{Horbach1999}%
  \BibitemOpen
  \bibfield  {author} {\bibinfo {author} {\bibfnamefont {J.}~\bibnamefont
  {{Horbach}}}\ and\ \bibinfo {author} {\bibfnamefont {W.}~\bibnamefont
  {{Kob}}},\ }\href@noop {} {\bibfield  {journal} {\bibinfo  {journal} {Phys.\
  Rev.~B}\ }\textbf {\bibinfo {volume} {60}},\ \bibinfo {pages} {3169}
  (\bibinfo {year} {1999})}\BibitemShut {NoStop}%
\bibitem [{\citenamefont {{Peckham}}(1967)}]{Peckham1966}%
  \BibitemOpen
  \bibfield  {author} {\bibinfo {author} {\bibfnamefont {G.}~\bibnamefont
  {{Peckham}}},\ }\href@noop {} {\bibfield  {journal} {\bibinfo  {journal}
  {Proc.\ Phys.\ Soc.\ London}\ }\textbf {\bibinfo {volume} {90}},\ \bibinfo
  {pages} {657} (\bibinfo {year} {1967})}\BibitemShut {NoStop}%
\bibitem [{\citenamefont {{Hazen}}(1976)}]{Hazen1976}%
  \BibitemOpen
  \bibfield  {author} {\bibinfo {author} {\bibfnamefont {R.~M.}\ \bibnamefont
  {{Hazen}}},\ }\href@noop {} {\bibfield  {journal} {\bibinfo  {journal} {Am.\
  Mineral.}\ }\textbf {\bibinfo {volume} {61}},\ \bibinfo {pages} {266}
  (\bibinfo {year} {1976})}\BibitemShut {NoStop}%
\bibitem [{\citenamefont {{Oganov}}, \citenamefont {{Gillan}},\ and\
  \citenamefont {{Price}}(2003)}]{Oganov2003}%
  \BibitemOpen
  \bibfield  {author} {\bibinfo {author} {\bibfnamefont {A.~R.}\ \bibnamefont
  {{Oganov}}}, \bibinfo {author} {\bibfnamefont {M.~J.}\ \bibnamefont
  {{Gillan}}}, \ and\ \bibinfo {author} {\bibfnamefont {G.~D.}\ \bibnamefont
  {{Price}}},\ }\href@noop {} {\bibfield  {journal} {\bibinfo  {journal}
  {J.~Chem.\ Phys.}\ }\textbf {\bibinfo {volume} {118}},\ \bibinfo {pages}
  {10174} (\bibinfo {year} {2003})}\BibitemShut {NoStop}%
\bibitem [{\citenamefont {{Tangney}}\ and\ \citenamefont
  {{Scandalo}}(2003)}]{Tangney2003}%
  \BibitemOpen
  \bibfield  {author} {\bibinfo {author} {\bibfnamefont {P.}~\bibnamefont
  {{Tangney}}}\ and\ \bibinfo {author} {\bibfnamefont {S.}~\bibnamefont
  {{Scandalo}}},\ }\href@noop {} {\bibfield  {journal} {\bibinfo  {journal}
  {J.~Chem.\ Phys.}\ }\textbf {\bibinfo {volume} {119}},\ \bibinfo {pages}
  {9673} (\bibinfo {year} {2003})}\BibitemShut {NoStop}%
\bibitem [{\citenamefont {{Rowley}}\ \emph {et~al.}(1998)\citenamefont
  {{Rowley}}, \citenamefont {{Jemmer}}, \citenamefont {{Wilson}},\ and\
  \citenamefont {{Madden}}}]{itapdb:Rowley1998}%
  \BibitemOpen
  \bibfield  {author} {\bibinfo {author} {\bibfnamefont {A.~J.}\ \bibnamefont
  {{Rowley}}}, \bibinfo {author} {\bibfnamefont {P.}~\bibnamefont {{Jemmer}}},
  \bibinfo {author} {\bibfnamefont {M.}~\bibnamefont {{Wilson}}}, \ and\
  \bibinfo {author} {\bibfnamefont {P.~A.}\ \bibnamefont {{Madden}}},\
  }\href@noop {} {\bibfield  {journal} {\bibinfo  {journal} {J.~Chem.\ Phys.}\
  }\textbf {\bibinfo {volume} {108}},\ \bibinfo {pages} {10209} (\bibinfo
  {year} {1998})}\BibitemShut {NoStop}%
\bibitem [{\citenamefont {{Fincham}}(1994)}]{itapdb:Fincham1994}%
  \BibitemOpen
  \bibfield  {author} {\bibinfo {author} {\bibfnamefont {D.}~\bibnamefont
  {{Fincham}}},\ }\href@noop {} {\bibfield  {journal} {\bibinfo  {journal}
  {Mol.\ Sim.}\ }\textbf {\bibinfo {volume} {13}},\ \bibinfo {pages} {1}
  (\bibinfo {year} {1994})}\BibitemShut {NoStop}%
\bibitem [{\citenamefont {{Kresse}}\ and\ \citenamefont
  {{Hafner}}(1993)}]{Kresse1993}%
  \BibitemOpen
  \bibfield  {author} {\bibinfo {author} {\bibfnamefont {G.}~\bibnamefont
  {{Kresse}}}\ and\ \bibinfo {author} {\bibfnamefont {J.}~\bibnamefont
  {{Hafner}}},\ }\href@noop {} {\bibfield  {journal} {\bibinfo  {journal}
  {Phys.\ Rev.~B}\ }\textbf {\bibinfo {volume} {47}},\ \bibinfo {pages} {558}
  (\bibinfo {year} {1993})}\BibitemShut {NoStop}%
\bibitem [{\citenamefont {{Kresse}}\ and\ \citenamefont
  {{Furthmüller}}(1996)}]{Kresse1996}%
  \BibitemOpen
  \bibfield  {author} {\bibinfo {author} {\bibfnamefont {G.}~\bibnamefont
  {{Kresse}}}\ and\ \bibinfo {author} {\bibfnamefont {J.}~\bibnamefont
  {{Furthmüller}}},\ }\href@noop {} {\bibfield  {journal} {\bibinfo  {journal}
  {Phys.\ Rev.~B}\ }\textbf {\bibinfo {volume} {54}},\ \bibinfo {pages} {11169}
  (\bibinfo {year} {1996})}\BibitemShut {NoStop}%
\bibitem [{\citenamefont {{Corona}}\ \emph {et~al.}(1987)\citenamefont
  {{Corona}}, \citenamefont {{Marchesi}}, \citenamefont {{Martini}},\ and\
  \citenamefont {{Ridella}}}]{itapdb:Corona1987}%
  \BibitemOpen
  \bibfield  {author} {\bibinfo {author} {\bibfnamefont {A.}~\bibnamefont
  {{Corona}}}, \bibinfo {author} {\bibfnamefont {M.}~\bibnamefont
  {{Marchesi}}}, \bibinfo {author} {\bibfnamefont {C.}~\bibnamefont
  {{Martini}}}, \ and\ \bibinfo {author} {\bibfnamefont {S.}~\bibnamefont
  {{Ridella}}},\ }\href@noop {} {\bibfield  {journal} {\bibinfo  {journal} {ACM
  Trans. Math. Softw.}\ }\textbf {\bibinfo {volume} {13}},\ \bibinfo {pages}
  {262} (\bibinfo {year} {1987})}\BibitemShut {NoStop}%
\bibitem [{\citenamefont {{Powell}}(1965)}]{itapdb:Powell1965}%
  \BibitemOpen
  \bibfield  {author} {\bibinfo {author} {\bibfnamefont {M.~J.~D.}\
  \bibnamefont {{Powell}}},\ }\href@noop {} {\bibfield  {journal} {\bibinfo
  {journal} {Comp.\ J.}\ }\textbf {\bibinfo {volume} {7}},\ \bibinfo {pages}
  {303} (\bibinfo {year} {1965})}\BibitemShut {NoStop}%
\bibitem [{\citenamefont {{Gropp}}, \citenamefont {{Lusk}},\ and\ \citenamefont
  {{Skjellum}}(1999)}]{itapdb:Gropp1999}%
  \BibitemOpen
  \bibfield  {author} {\bibinfo {author} {\bibfnamefont {W.}~\bibnamefont
  {{Gropp}}}, \bibinfo {author} {\bibfnamefont {E.}~\bibnamefont {{Lusk}}}, \
  and\ \bibinfo {author} {\bibfnamefont {A.}~\bibnamefont {{Skjellum}}},\
  }\href@noop {} {\emph {\bibinfo {title} {Using MPI - 2nd edn}}}\ (\bibinfo
  {publisher} {Cambridge, MA: MIT Press},\ \bibinfo {year} {1999})\BibitemShut
  {NoStop}%
\bibitem [{\citenamefont {{Kresse}}\ and\ \citenamefont
  {{Joubert}}(1999)}]{itapdb:Kresse1999}%
  \BibitemOpen
  \bibfield  {author} {\bibinfo {author} {\bibfnamefont {G.}~\bibnamefont
  {{Kresse}}}\ and\ \bibinfo {author} {\bibfnamefont {D.}~\bibnamefont
  {{Joubert}}},\ }\href@noop {} {\bibfield  {journal} {\bibinfo  {journal}
  {Phys.\ Rev.~B}\ }\textbf {\bibinfo {volume} {59}},\ \bibinfo {pages} {1758}
  (\bibinfo {year} {1999})}\BibitemShut {NoStop}%
\bibitem [{\citenamefont {{Carr\'e}}\ \emph {et~al.}(2007)\citenamefont
  {{Carr\'e}}, \citenamefont {{Berthier}}, \citenamefont {{Horbach}},
  \citenamefont {{Ispas}},\ and\ \citenamefont {{Kob}}}]{Carre2007}%
  \BibitemOpen
  \bibfield  {author} {\bibinfo {author} {\bibfnamefont {A.}~\bibnamefont
  {{Carr\'e}}}, \bibinfo {author} {\bibfnamefont {L.}~\bibnamefont
  {{Berthier}}}, \bibinfo {author} {\bibfnamefont {J.}~\bibnamefont
  {{Horbach}}}, \bibinfo {author} {\bibfnamefont {S.}~\bibnamefont {{Ispas}}},
  \ and\ \bibinfo {author} {\bibfnamefont {W.}~\bibnamefont {{Kob}}},\
  }\href@noop {} {\bibfield  {journal} {\bibinfo  {journal} {J.~Chem.\ Phys.}\
  }\textbf {\bibinfo {volume} {127}},\ \bibinfo {pages} {114512} (\bibinfo
  {year} {2007})}\BibitemShut {NoStop}%
\bibitem [{\citenamefont {{Vollmayr}}, \citenamefont {{Kob}},\ and\
  \citenamefont {{Binder}}(1996)}]{Vollmayr1996}%
  \BibitemOpen
  \bibfield  {author} {\bibinfo {author} {\bibfnamefont {K.}~\bibnamefont
  {{Vollmayr}}}, \bibinfo {author} {\bibfnamefont {W.}~\bibnamefont {{Kob}}}, \
  and\ \bibinfo {author} {\bibfnamefont {K.}~\bibnamefont {{Binder}}},\
  }\href@noop {} {\bibfield  {journal} {\bibinfo  {journal} {Phys.\ Rev.~B}\
  }\textbf {\bibinfo {volume} {54}} (\bibinfo {year} {1996})}\BibitemShut
  {NoStop}%
\bibitem [{\citenamefont {{Rog}}\ \emph {et~al.}(2003)\citenamefont {{Rog}},
  \citenamefont {{Murzyn}}, \citenamefont {{Hinsen}},\ and\ \citenamefont
  {{Kneller}}}]{Rog2003}%
  \BibitemOpen
  \bibfield  {author} {\bibinfo {author} {\bibfnamefont {T.}~\bibnamefont
  {{Rog}}}, \bibinfo {author} {\bibfnamefont {K.}~\bibnamefont {{Murzyn}}},
  \bibinfo {author} {\bibfnamefont {K.}~\bibnamefont {{Hinsen}}}, \ and\
  \bibinfo {author} {\bibfnamefont {G.~R.}\ \bibnamefont {{Kneller}}},\
  }\href@noop {} {\bibfield  {journal} {\bibinfo  {journal} {J.~Comput.\
  Chem.}\ }\textbf {\bibinfo {volume} {24}},\ \bibinfo {pages} {657} (\bibinfo
  {year} {2003})}\BibitemShut {NoStop}%
\bibitem [{\citenamefont {{Gibbs}}\ \emph {et~al.}(2009)\citenamefont
  {{Gibbs}}, \citenamefont {{Wallace}}, \citenamefont {{Cox}}, \citenamefont
  {{Downs}}, \citenamefont {{Ross}},\ and\ \citenamefont {Rosso}}]{Gibbs2009}%
  \BibitemOpen
  \bibfield  {author} {\bibinfo {author} {\bibfnamefont {G.~V.}\ \bibnamefont
  {{Gibbs}}}, \bibinfo {author} {\bibfnamefont {A.~F.}\ \bibnamefont
  {{Wallace}}}, \bibinfo {author} {\bibfnamefont {D.~F.}\ \bibnamefont
  {{Cox}}}, \bibinfo {author} {\bibfnamefont {R.~T.}\ \bibnamefont {{Downs}}},
  \bibinfo {author} {\bibfnamefont {N.~L.}\ \bibnamefont {{Ross}}}, \ and\
  \bibinfo {author} {\bibfnamefont {K.~M.}\ \bibnamefont {Rosso}},\ }\href@noop
  {} {\bibfield  {journal} {\bibinfo  {journal} {ammin}\ }\textbf {\bibinfo
  {volume} {94}},\ \bibinfo {pages} {1085} (\bibinfo {year}
  {2009})}\BibitemShut {NoStop}%
\bibitem [{\citenamefont {{Gibbs}}\ \emph {et~al.}(2006)\citenamefont
  {{Gibbs}}, \citenamefont {{Jayatilaka}}, \citenamefont {{Spackman}},
  \citenamefont {{Cox}},\ and\ \citenamefont {Rosso}}]{Gibbs2006}%
  \BibitemOpen
  \bibfield  {author} {\bibinfo {author} {\bibfnamefont {G.~V.}\ \bibnamefont
  {{Gibbs}}}, \bibinfo {author} {\bibfnamefont {D.}~\bibnamefont
  {{Jayatilaka}}}, \bibinfo {author} {\bibfnamefont {M.~A.}\ \bibnamefont
  {{Spackman}}}, \bibinfo {author} {\bibfnamefont {D.~F.}\ \bibnamefont
  {{Cox}}}, \ and\ \bibinfo {author} {\bibfnamefont {K.~M.}\ \bibnamefont
  {Rosso}},\ }\href@noop {} {\bibfield  {journal} {\bibinfo  {journal} {J.
  Phys. Chem. A}\ }\textbf {\bibinfo {volume} {110}},\ \bibinfo {pages} {12678}
  (\bibinfo {year} {2006})}\BibitemShut {NoStop}%
\bibitem [{\citenamefont {{Deserno}}\ and\ \citenamefont {{Holm}}(1998)}]{P3M}%
  \BibitemOpen
  \bibfield  {author} {\bibinfo {author} {\bibfnamefont {M.}~\bibnamefont
  {{Deserno}}}\ and\ \bibinfo {author} {\bibfnamefont {C.}~\bibnamefont
  {{Holm}}},\ }\href@noop {} {\bibfield  {journal} {\bibinfo  {journal}
  {J.~Chem.\ Phys.}\ }\textbf {\bibinfo {volume} {109}},\ \bibinfo {pages}
  {7678} (\bibinfo {year} {1998})}\BibitemShut {NoStop}%
\bibitem [{\citenamefont {{Limbach}}\ \emph {et~al.}(2006)\citenamefont
  {{Limbach}}, \citenamefont {{Arnold}}, \citenamefont {{Mann}},\ and\
  \citenamefont {{Holm}}}]{Espresso}%
  \BibitemOpen
  \bibfield  {author} {\bibinfo {author} {\bibfnamefont {H.-J.}\ \bibnamefont
  {{Limbach}}}, \bibinfo {author} {\bibfnamefont {A.}~\bibnamefont {{Arnold}}},
  \bibinfo {author} {\bibfnamefont {B.~A.}\ \bibnamefont {{Mann}}}, \ and\
  \bibinfo {author} {\bibfnamefont {C.}~\bibnamefont {{Holm}}},\ }\href@noop {}
  {\bibfield  {journal} {\bibinfo  {journal} {Comput. Phys. Commun.}\ }\textbf
  {\bibinfo {volume} {174}},\ \bibinfo {pages} {704} (\bibinfo {year}
  {2006})},\ \bibinfo {note} {{\tt
  http://www.http://espressomd.org}}\BibitemShut {NoStop}%
\bibitem [{\citenamefont {{Ghose}}\ \emph {et~al.}(2006)\citenamefont
  {{Ghose}}, \citenamefont {{Krisch}}, \citenamefont {{Oganov}}, \citenamefont
  {{Beraud}}, \citenamefont {{Bosak}}, \citenamefont {{Gulve}}, \citenamefont
  {{Seelaboyina}}, \citenamefont {{Yang}},\ and\ \citenamefont
  {{Saxena}}}]{Ghose2006}%
  \BibitemOpen
  \bibfield  {author} {\bibinfo {author} {\bibfnamefont {S.}~\bibnamefont
  {{Ghose}}}, \bibinfo {author} {\bibfnamefont {M.}~\bibnamefont {{Krisch}}},
  \bibinfo {author} {\bibfnamefont {A.~R.}\ \bibnamefont {{Oganov}}}, \bibinfo
  {author} {\bibfnamefont {A.}~\bibnamefont {{Beraud}}}, \bibinfo {author}
  {\bibfnamefont {A.}~\bibnamefont {{Bosak}}}, \bibinfo {author} {\bibfnamefont
  {R.}~\bibnamefont {{Gulve}}}, \bibinfo {author} {\bibfnamefont
  {R.}~\bibnamefont {{Seelaboyina}}}, \bibinfo {author} {\bibfnamefont
  {H.}~\bibnamefont {{Yang}}}, \ and\ \bibinfo {author} {\bibfnamefont {S.~K.}\
  \bibnamefont {{Saxena}}},\ }\href@noop {} {\bibfield  {journal} {\bibinfo
  {journal} {Phys.\ Rev.\ Lett.}\ }\textbf {\bibinfo {volume} {96}},\ \bibinfo
  {pages} {035507} (\bibinfo {year} {2006})}\BibitemShut {NoStop}%
\bibitem [{\citenamefont {{Bosak}}\ and\ \citenamefont
  {{Krisch}}(2005)}]{Bosak2005}%
  \BibitemOpen
  \bibfield  {author} {\bibinfo {author} {\bibfnamefont {A.}~\bibnamefont
  {{Bosak}}}\ and\ \bibinfo {author} {\bibfnamefont {M.}~\bibnamefont
  {{Krisch}}},\ }\href@noop {} {\bibfield  {journal} {\bibinfo  {journal}
  {Phys.\ Rev.~B}\ }\textbf {\bibinfo {volume} {72}},\ \bibinfo {pages}
  {224305} (\bibinfo {year} {2005})}\BibitemShut {NoStop}%
\bibitem [{\citenamefont {{Hocker}}\ \emph {et~al.}()\citenamefont {{Hocker}},
  \citenamefont {{Beck}}, \citenamefont {{Roth}}, \citenamefont {{Schmauder}},\
  and\ \citenamefont {{Trebin}}}]{Hocker2011}%
  \BibitemOpen
  \bibfield  {author} {\bibinfo {author} {\bibfnamefont {S.}~\bibnamefont
  {{Hocker}}}, \bibinfo {author} {\bibfnamefont {P.}~\bibnamefont {{Beck}}},
  \bibinfo {author} {\bibfnamefont {J.}~\bibnamefont {{Roth}}}, \bibinfo
  {author} {\bibfnamefont {S.}~\bibnamefont {{Schmauder}}}, \ and\ \bibinfo
  {author} {\bibfnamefont {H.-R.}\ \bibnamefont {{Trebin}}},\ }\href@noop {} {\
  \bibinfo {series} {(unpublished)}}\BibitemShut {NoStop}%
\end{thebibliography}
\end{document}